# Ultimate limit to human longevity

Byung Mook Weon

*LG.Philips Displays, 184, Gongdan1-dong, Gumi-city, GyungBuk, 730-702, Korea*

*Author for correspondence (bmw@lgphilips-displays.com)*

Are there limits to human longevity? We suggest a new demographic model to describe human demographic trajectories. Specifically, the model mathematically defines the limits of longevity. Through the demographic analysis of trends for Sweden (between 1751 and 2002), Switzerland (between 1876 and 2002) and Japan (between 1950 and 1999), which are the longest-lived countries, we would like to demonstrate whether or not there is the ultimate limit to longevity. We analyse the trends of new demographic indicators, the characteristic life and the shape parameter, and calculate the mathematical limits of longevity. We find out the surprising phenomenon that the mathematical limits of longevity decrease as the longevity tendency increases in recent decades. These paradoxical trends will be explained by the complementarity of longevity, which is attributable to the nature of biological systems for longevity. According to the regression analysis, the ultimate limit for humans is estimated to be approximately 124 years, which may be considered to be the biological limit for humans.

## 1. Introduction

A fundamental question in ageing research is whether humans possess ultimate limits to longevity (Wilmoth *et al.* 2000). Such a question has yet to be resolved, although more and more scientists are coming to believe that human longevity may be on the increase (Vaupel *et al.* 1998; Wilmoth *et al.* 2000; Long-lived bet 2001; Oeppen & Vaupel 2002). However, Hayflick (2000) pointed out, "If we are to increase human life expectancy beyond the fifteen-year limit that would result if today's leading causes of death were resolved, more attention must be paid to basic research on ageing". According to Vaupel *et al.* (2003), "Since 1840, record life expectancy has increased by 2.5 years per decade, but will this march to longevity continue for many more decades?" If there are increases and/or limitations in the trends of human longevity at present, then how should we analyse it? The conventional demographic model and methodology in the field of demography and biology seem to be not fully successful to identify whether there are limits to human longevity (Wilmoth 1997).

## 2. Material

For analysis, we used the survival data for Sweden (between 1751 and 2002), Switzerland (between 1876 and 2002), and Japan (between 1950 and 1999), which are the longest-lived countries, from the period life tables (for all sexes, 1x1) taken from the Human Mortality Database (www.mortality.org). The all period, which are available for each country in the Human Mortality Database, were chosen. The survival probability ($S$) is expressed as a fraction ($l_x/l_o$) of the number of survivors ($l_x$) out of 100,000 persons ($l_o$) in the original life tables.

**3. Model**

In order to describe human demographic trajectories, a new demographic model has been found by Weon for the first time to our knowledge. We would like to call this new model the "Weon model" hereafter, as the Gompertz model was formulated by Benjamin Gompertz and thus it bears his name. The Weon model is derived from the Weibull model (Weibull 1951) with an assumption that the shape parameter is a function of age. In the Weibull model, the shape parameter is constant with age (Nelson 1990). The age-dependent shape parameter enables us to model the demographic (survival and mortality) functions, which are expressed as follows,

$$S(t) = \exp(-(t/\alpha)^{\beta(t)})$$

$$\mu(t) = (t/\alpha)^{\beta(t)} \times [\frac{\beta(t)}{t} + \ln(t/\alpha) \times \frac{d\beta(t)}{dt}]$$

where $S(t)$ is the survival function, indicating the probability that an individual is still alive at age $t$ and $\mu(t)$ is the mortality function, indicating the probability density at age conditional on survival to that age, in which $\alpha$ denotes the characteristic life ($t = \alpha$ when $S(t) = \exp(-1) \approx 36.79\%$) and $\beta(t)$ denotes the shape parameter as a function of age. The original idea was obtained as follows: typical human survival curves show i) a rapid decrease in survival in the first few years of life and then ii) a relatively steady decrease and then an abrupt decrease near death. Interestingly, the former behaviour resembles the Weibull survival function with $\beta < 1$ and the latter behaviour seems to follow the case of $\beta >> 1$. With this in mind, it could be assumed that shape parameter is a function of age.

The Weon model is completely different with the Weibull model in the age dependence of the shape parameter. We could evaluate the age dependence of the shape parameter to determine an adequate mathematical expression of the shape parameter, after determination of the characteristic life graphically in the survival curve. Conveniently, the value of the characteristic life is always found at the duration for the survival to be '$\exp(-1)$'; this is known as the characteristic life. This feature gives the advantage of looking for the value of $\alpha$ simply by graphical analysis of the survival curve. In turn, with the observed value of $\alpha$, we can plot the shape parameter as a function of age by the mathematical equivalence of '$\beta(t) = \ln(-\ln S(t)) / \ln(t/\alpha)$'. If $\beta(t)$ is not constant with age, this obviously implies that '$\beta(t)$ is a function of age'. In empirical practice, we could successfully use a polynomial expression for modelling the shape parameter as a function of age as follows: $\beta(t) = \beta_0 + \beta_1 t + \beta_2 t^2 + ...$, where the associated coefficients could be determined by a regression analysis in the plot of shape parameter curve. And thus, the derivative is obtained as follows: $d\beta(t)/dt = \beta_1 + 2\beta_2 t + ...$, which indicates again that the shape parameter for humans is a function of age. If $\beta(t)$ (except for the mathematical singularity or trace of $\alpha$) can be expressed by an adequate mathematical function, the survival and mortality functions can be calculated by the mathematically expressed $\beta(t)$.

As we know, no mathematical model has been suggested that can perfectly approximate the development of the mortality rate over the total life span (Kowald 1999). In practice for the Weon model, a linear expression for $\beta(t)$ is roughly appropriate for ages before $\alpha$ and a quadratic expression is appropriate for ages after $\alpha$. With these mathematical expressions of the shape parameter, we could solve the conventional problems to describe the demographic trajectories. The fundamental demographic model is the Gompertz model (Gompertz 1825), in which the human mortality rate increases roughly exponentially with increasing age at senescence. However, the mortality rate does not increase according to the Gompertz model at the highest ages (Vaupel 1997; Thatcher *et al.* 1998; Robine & Vaupel 2001; 2002; Yi & Vaupel 2003), and this deviation from the Gompertz model is a great puzzle to demographers, biologists and gerontologists (Vaupel 1997). Furthermore, it is still not certain whether the mortality trajectories level or decrease at the highest ages (Vaupel 1997; Vaupel et al. 1998; Thatcher 1999; Lynch & Brown 2001; Robine & Vaupel 2001; 2002; Helfand & Inouye 2002). By the way, the Weon model through the quadratic expression for $\beta(t)$ predicts that the mortality rate inevitably decreases after a plateau and ultimately approach zero. Furthermore, through the approximate relationship of '$\ln \mu(t) \propto \beta(t)$' after adulthood (~30-80) by the linear expression for $\beta(t)$, the Weon model approximates the Gompertz model when '$\beta(t) \propto t$'. Particularly, the mortality rate would deviate from the Gompertz model when $\beta(t)$ has a non-linear behaviour. It is therefore possible that $\beta(t)$ is a measure of the deviation from the Gompertz model at the highest ages. The Weon model can therefore generalize the Gompertz model as well as the Weibull model: That is, the Gompertz model is a special case of a linear expression for $\beta(t)$ and the Weibull model is a special case of a constant shape parameter.

According to the above mortality function, the Weon model suggests a simple mathematical definition of limits of longevity as follows: In principle, the mortality function should be mathematically positive ( $\mu(t) \geq 0$ ). Therefore, the mathematical criterion for limits of longevity, which is able to be determined by the mortality trajectories in nature, can be given by,

$$\frac{d\beta(t)}{dt} \geq -\frac{\beta(t)}{t\ln(t/\alpha)}$$

In fact, it is possible that the survival function calculated through modeling the shape parameter for the highest ages is not zero, although it has extremely low values at the highest ages, but that the mortality function for the highest ages can reach zero at the mathematical limit of longevity. In this case, the decrease rate of the survival function ( $-dS(t)/dt$ ; this term means the probability density function, and the minus indicates the decrease) should be zero at the mathematical limit of longevity. Therefore, the mathematical limit of longevity can be simply defined as "$-dS(t)/dt = 0$ or $\mu(t) = 0$". It should be noted that the survival and mortality functions are linked to the mathematical relationship of "$\mu(t) = -dS(t)/dt \times 1/S(t)$". In practice, we can identify the mathematical limit of longevity at the moment that the survival trajectory levels off, or the mortality trajectory becomes zero.

**4. Results**

Evidently, we could see the age dependence of the shape parameter for humans, through the trajectories of shape parameter plotted for Sweden, Switzerland and Japan in Fig. 1 (a)~(c). On the other hand, $\beta(t)$ mathematically approaches infinity as the age $t$ approaches the value of $\alpha$ or the denominator '$\ln(t/\alpha)$' approaches zero. This feature leaves the 'trace of $\alpha$' in the plot of $\beta(t)$, thus we could observe variations of $\beta(t)$ and $\alpha$ at once, as shown in Fig. 1 (a)~(c).

We carried out the regression analysis for ages after $\alpha$ for Sweden (between 1751 and 2002) in Table S1, for Switzerland (between 1876 and 2002) in Table S2 and for Japan (between 1950 and 1999) in Table S3 (in Supplementary Table), using the quadratic expression for $\beta(t)$. Specifically, the quadratic coefficient is important because it determines the mathematical limit of longevity in consequence of the quadratic expression of $\beta(t)$ for the highest ages. With this in mind, we would like to demonstrate the historical trends of the characteristic life and quadratic coefficient over time. First, the characteristic life has increased constantly for more than a century in Fig. 2 (a), which is consistent with other literature, although the rates of increase are slightly different by country (Vaupel *et al.* 1998; Wilmoth *et al.* 2000; Oeppen & Vaupel 2002; Vaupel *et al.* 2003). The characteristic life has increased by approximately 1.48 years per decade for Sweden (between 1751 and 2002), 2.08 years per decade for Switzerland (between 1876 and 2002) and 2.77 years for Japan (between 1950 and 1999), respectively. Second, it could be seen in Fig. 2 (b) that the quadratic coefficient of the shape parameter, which indicates the decrease (or bending down) of the shape parameter at the highest ages, has similarly increased over time for the countries. Particularly, significant increases have taken place since 1970s.

Yet to our surprise, it is likely to be obvious that the mathematical limit of longevity tends to decrease as the quadratic coefficient increases in Fig. 3. In other words, the trend of the mathematical limit runs in clear contradiction to the trend of the characteristic life. Why seems to be there such a contradiction? To answer this question, we need to reconsider what the origin of the age-dependent shape parameter is. In principle for the highest value of the survival trajectory or for longevity at all times, the shape parameter should be variable according to the characteristic life: "for longevity, $\beta(t)$ tends to increase at $t < \alpha$ but it tends to decrease at $t > \alpha$." This is attributable to the nature of biological systems to strive to survive healthier and longer robustly against intrinsic defects and circumstances (Kirkwood & Austad 2000; Ball 2002). This tendency seems to be obvious in the Fig. 1 (a)~(c). Empirically, we already know that the quadratic coefficient indicates the decrease (or bending down) of $\beta(t)$ at $t > \alpha$. The characteristic life tends to increase over time as well as the quadratic coefficient in Fig. 2 (a) and (b). By contraries, the mathematical limit of longevity tends to decrease with the quadratic coefficient in Fig. 3. Therefore, we come to the conclusion that the mathematical limit of longevity decreases as the longevity tendency (the characteristic life) increases. We call this phenomenon the "complementarity of longevity". That is, the longevity tendency bends down the shape parameter after the characteristic life, which simultaneously is the reason of decreasing the mathematical limit of longevity. The complementarity caused by the longevity tendency is attributable to the nature of biological systems to strive to survive healthier and longer robustly against intrinsic defects and circumstances. It is the age dependence of the shape parameter that is likely to be governed by the complementarity of longevity.

## 5. Discussion

Where is the ultimate limit to human longevity? Finally, we are able to see that the trend line (if denoted as "$\omega$") empirically fitted by an exponential model of the mathematical limits with the quadratic coefficient seems to be "$\omega \approx 124.0 + 183.1 \times \exp(\beta_2 / 0.0004)$" ($r^2 = 0.9395$) in Fig. 3. Note that $\beta_2$ is negative.

In this case, the ultimate value of the mathematical limits as '$\beta_2 \to -\infty$', assuming that the longevity tendency increases extremely, leading to that the bending down of the shape parameter is maximized, can be predicted to be approximately 124 years totally for Sweden, Switzerland and Japan. Interestingly, this value is approximate to the world record for human longevity of Jeanne Calment, who lived to the ripe old age of 122 years and 164 days (122.45 years) (Vallin & Meslé 2001).

Considering the scattering of the data points, the results suggest that the human longevity should be ultimately limited around 120 years, which may be associated with the biological limits (Hayflick 1996; 1998; 2000; Carnes *et al.* 2003) and the demographic evidence in history (Shapin & Martyn 2000). The human life span has remained unchanged for the past 100,000 years at about 125 years (Hayflick 1996; 1998; 2000). Although the life expectancy is likely to continuously increase, the ultimate limit to longevity is not likely to be broken. It is probably impossible to surpass the barrier without intrinsic (or biological) changes of human body.

Moreover, the significant increase of longevity since 1970s (Fig. 2 (b)) is likely to be associated with the improvement of medical care, for example, an introduction of heart transplant. Today, in developed countries more than 75% of all deaths now occur in those over the age of 75 (Hayflick 2000). Of the total increase of longevity, 72.5% is attributable to a decline in mortality above age 70 (Wilmoth *et al.* 2000). Maybe further improvement of medical care will be effective to continuously increase the longevity (life expectancy, or characteristic life) in a population, although it is difficult to surpass the ultimate limit around 120 years. If so, we may expect that the older persons will continue to accumulate between 80 and 120 years. Then, we will need a policy and expenditures to take care of the older persons to be piled up in the forthcoming future (Friedland 1998; Lubitz *et al.* 2003).

**Acknowledgements**

The author thanks to the Human Mortality Database (Dr. John R. Wilmoth, as a director, in The University of California, Berkeley and Dr. Vladimir Shkolnikov, as a co-director, in The Max Planck Institute for Demographic Research) for allowing anyone to access the demographic data for research.

## Supplementary Tables

Table S1. Regression analysis for ages after characteristic life for Sweden between 1751 and 2002, and calculated characteristic life and mathematical limit.

Table S2. Regression analysis for ages after characteristic life for Switzerland between 1876 and 2002, and calculated characteristic life and mathematical limit.

Table S3. Regression analysis for ages after characteristic life for Japan between 1950 and 1999, and calculated characteristic life and mathematical limit.

Table S1.xls Table S2.xls Table S3.xls

## Supplementary Information

Definitions and relationships of the Weon model are described at the following file.

Supplementary Info.doc

**Figure legends**

Fig. 1. Trajectories of shape parameter as a function of age: (a) for Sweden (between 1751 and 2002), (b) Switzerland (between 1876 and 2002), and (c) Japan (between 1950 and 1999).

Fig. 2. Trends of longevity: (a) characteristic life and (b) quadratic coefficient of shape parameter after characteristic life for Sweden (between 1751 and 2002), Switzerland (between 1876 and 2002), and Japan (between 1950 and 1999).

Fig. 3. Trends of limits to longevity with quadratic coefficient for Sweden (between 1751 and 2002), Switzerland (between 1876 and 2002), and Japan (between 1950 and 1999).

**Figures**

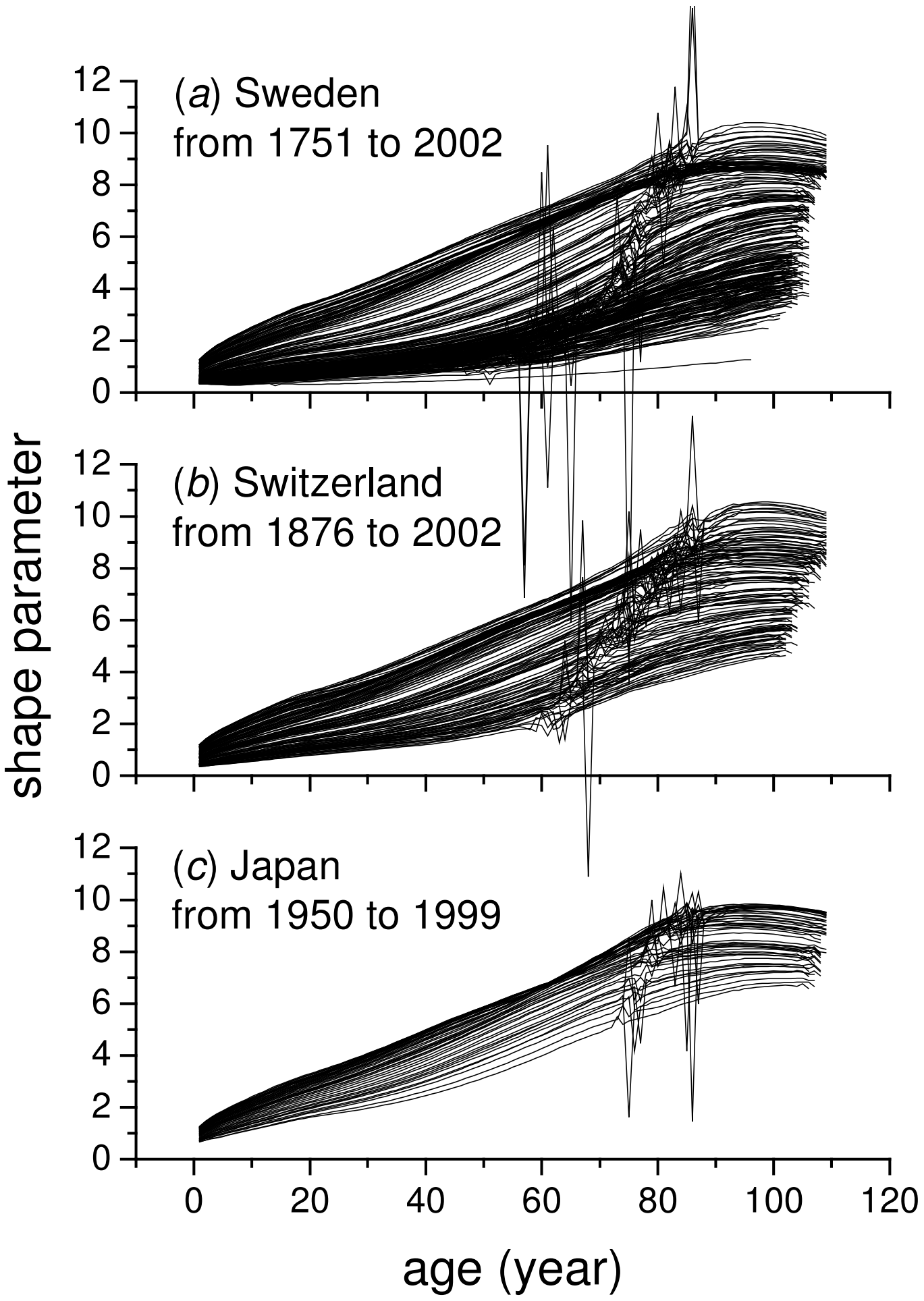

Figure 1.

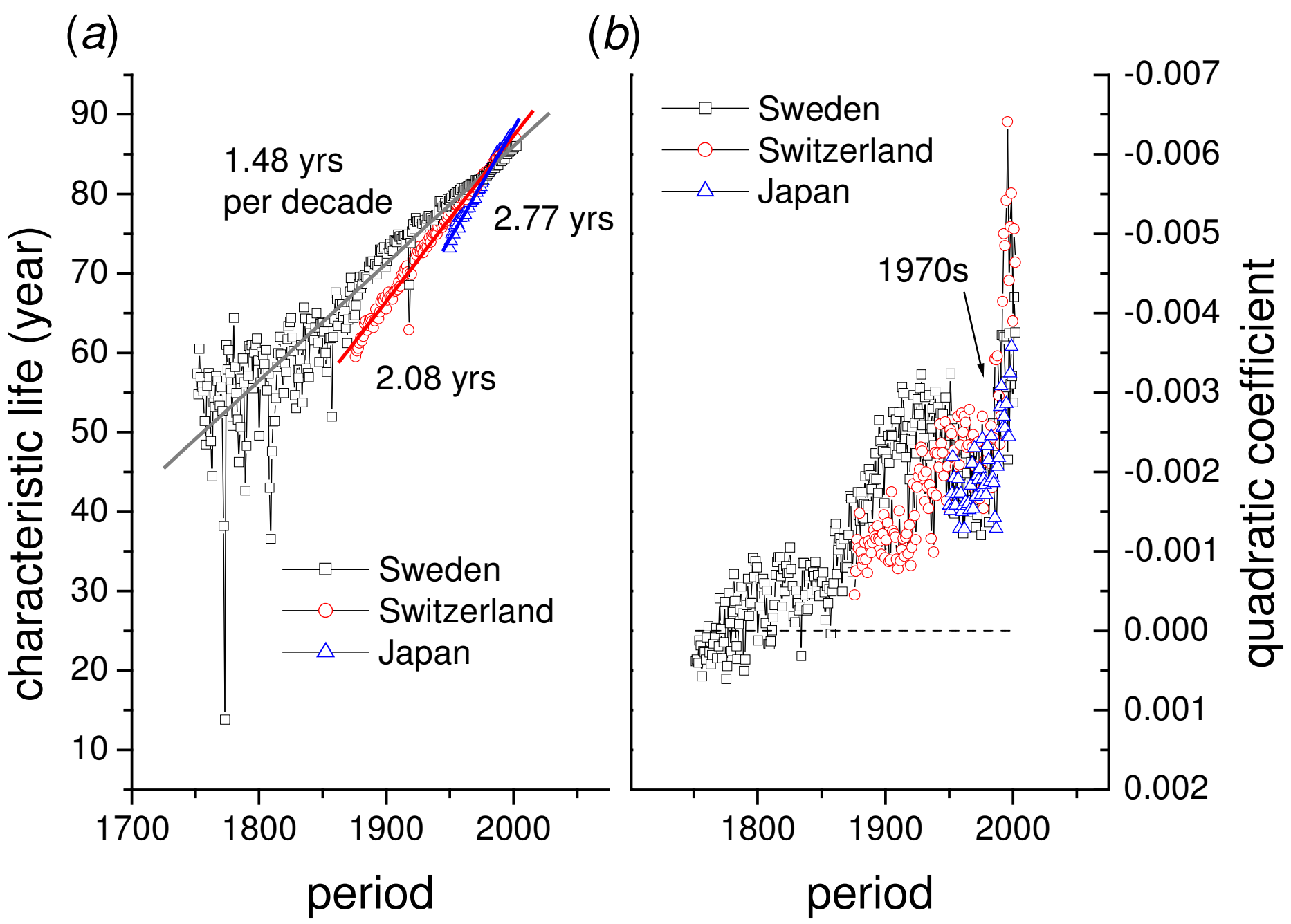

Figure 2.

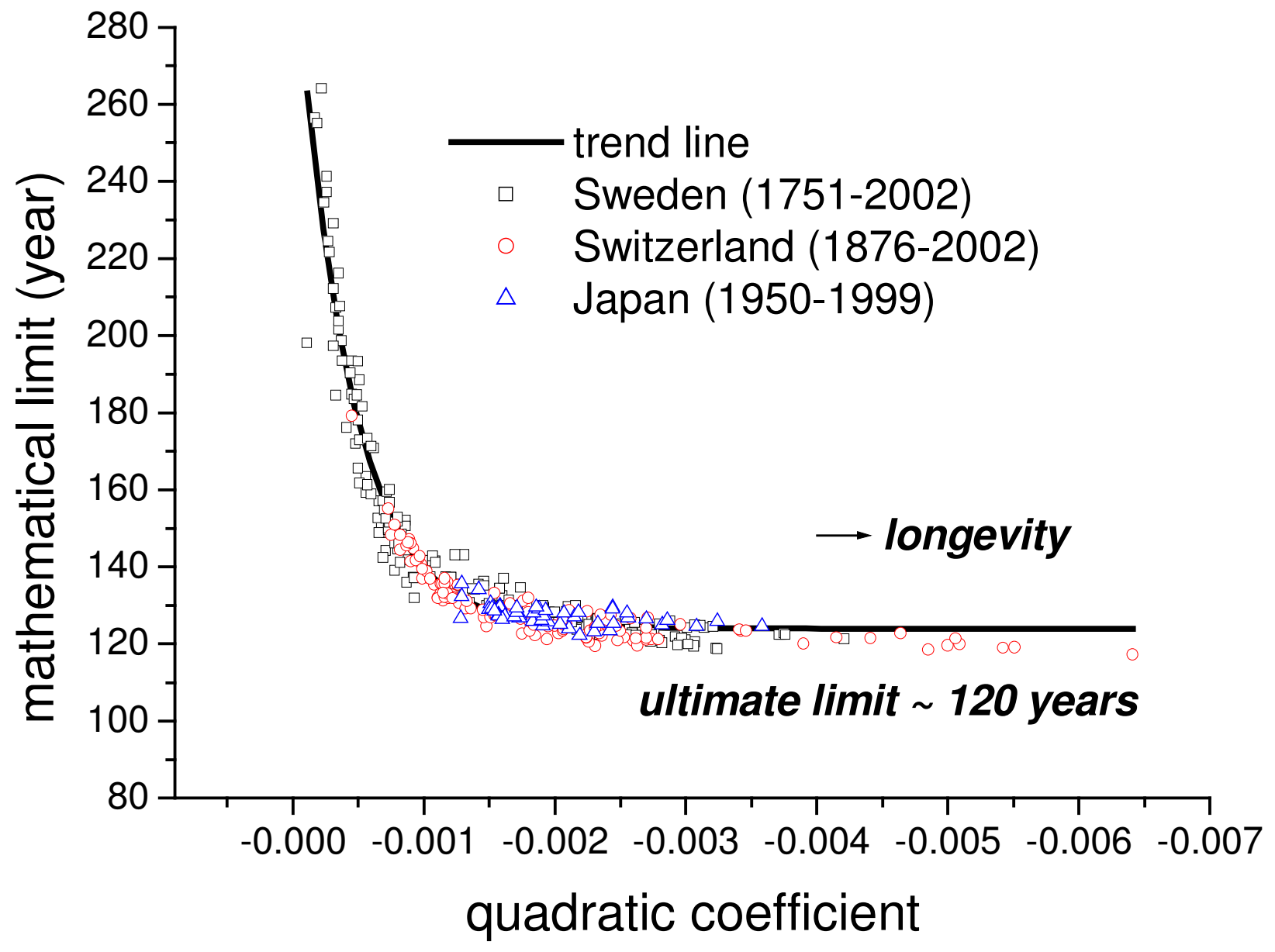

Figure 3.

# Supplementary Information

Byung Mook Weon
*LG.Philips Displays, 184, Gongdan1-dong, Gumi-city, GyungBuk, 730-702, Korea*
*Author for correspondence (bmw@lgphilips-displays.com)*

Supplementary information for "Ultimate limit to human longevity" submitted to *Proceedings B*.

## 1. Weon model: Empirical descriptive model for humans

The Weon model is a modified Weibull model with an age-dependent shape parameter as follows. The age dependence of shape parameter is attributable to dynamical aspects of biological system (humans).

1) Survival function: $S(t) = \exp(-(t/\alpha)^{\beta(t)})$

2) Shape parameter: $\beta(t) = \ln(-\ln S(t))/\ln(t/\alpha)$, which empirically can be expressed by '$\beta(t) = \beta_0 + \beta_1 t + \beta_2 t^2 + ...$' as a function of age.

3) Mortality function: $\mu(t) = (t/\alpha)^{\beta(t)} \times [\frac{\beta(t)}{t} + \ln(t/\alpha) \times \frac{d\beta(t)}{dt}]$

4) Density function: $f(t) = S(t) \times \mu(t)$

## 2. Statistical relationships

For statistical definitions, let $f(t)$ be the *probability density function* (*pdf*) describing the distribution of life spans in a population. The *cumulative density function* (*cdf*), $F(t)$, gives the probability that an individual dies before surpassing age $t$ (especially age $t$ is a *continuous random variable*). The *survival function*, $S(t)$, gives the complementary probability ($S(t) = 1 - F(t)$) that an individual is still alive at age $t$. The *mortality function*, $\mu(t)$, is defined as the ratio of the density and survival functions ($\mu(t) = f(t)/S(t)$). Thus, the mortality function gives the probability density at age $t$ conditional on survival to that age. $S(t)$, $\beta(t)$, $\mu(t)$ and $f(t)$ are variables as a function of age, and $\alpha$, $\beta_0$, $\beta_1$ and $\beta_2$ are constants.

1) Cumulative density function (*cdf*): $F(t) = 1 - S(t)$

2) Probability density function (*pdf*): $f(t) = \frac{dF(t)}{dt} = -\frac{dS(t)}{dt}$ and $f(t) = S(t) \times \mu(t)$

3) Survival function: $S(t) = 1 - F(t)$

4) Mortality function (instantaneous hazard function): $\mu(t) = \frac{f(t)}{S(t)} = -\frac{d\ln S(t)}{dt}$

## 3. Singular points

1) Characteristic life ($\alpha$)

Conveniently, the value of the characteristic life is always found at the duration for the survival to be '$S(t) = \exp(-1)$'; this is known as the characteristic life (figure A1 (a)). This feature gives the advantage of looking for the value of $\alpha$ simply by graphical analysis of the survival curve. The value of $\beta(t)$ mathematically approaches infinity as the age $t$ approaches the value of $\alpha$ or the denominator '$\ln(t/\alpha)$' approaches zero. This feature leaves the 'trace of $\alpha$' in the plot of $\beta(t)$, thus we can observe variations of $\beta(t)$ and $\alpha$ at once (figure A1 (b)). Empirically there seems to exist the maximum value (or the peak) of $f(t)$ approximately at $\alpha$ (figure A1 (c)).

i) $t = \alpha$ at $S(t) = \exp(-1)$

ii) $\beta(t) \to \infty$ at $t \to \alpha$. Traces of $\alpha$ can be shown in plot of $\beta(t)$.

iii) Empirically $\dfrac{df(t)}{dt} \approx 0$ at $t \approx \alpha$

2) Vertex of shape parameter ($\nu$)

It indicates a maximum value of shape parameter after characteristic life (figure A1 (b)): $\dfrac{d\beta(t)}{dt} = 0$ at $t = \nu$ where $\nu = -\dfrac{\beta_1}{2\beta_2}$.

3) Plateau of mortality ($\rho$)

It indicates a maximum value of mortality function after characteristic life (figure A1 (d)): $\dfrac{d\mu(t)}{dt} = 0$ at $t = \rho$.

4) Mathematical limit of longevity or shortly 'maximum longevity' ($\omega$)

Perhaps the most common notion of a limit in the study of human longevity is the *limited-life-span hypothesis*, which states that there exists some age ($\omega$) beyond which there can be no survivors. This hypothesis can be expressed by any one of the following three formulas: "$f(t) = 0$, $S(t) = 0$ or $\lim_{t \to \omega} \mu(t) = \infty$ ($t \geq \omega$)."

However, according to the Weon model, the survival function may be not zero, although it has extremely low values at the highest ages, while the mortality function can be zero at the maximum longevity (figure A1 (d)). The Weon model suggests that the maximum longevity can be defined as follows: "at $t = \omega$, $f(t) = 0$ and $\mu(t) = 0$, instead of $S(t) = 0$." Fundamentally, the decrease rate of the survival function with age ($-dS(t)/dt$; this term means the density function (*pdf*), $f(t)$, and the minus indicates the decrease) should be zero at the maximum longevity. Therefore, the maximum longevity can be simply defined as "$-dS(t)/dt = 0$ or $\mu(t) = 0$". In practice, we can identify the maximum longevity at the moment that the survival trajectory levels off, or the mortality trajectory becomes zero.

i) $f(t)=0$ or $-\frac{dS(t)}{dt}=0$ at $t=\omega$

ii) $\mu(t)=0$ or $\frac{d\beta(t)}{dt}=-\frac{\beta(t)}{t\times\ln(t/\alpha)}$ at $t=\omega$

**4. Indicators of longevity**

1) Characteristic life ($\alpha$) and vertex of shape parameter ($\nu$) are observable indicators of longevity in the trends of plot of shape parameter; there is empirically an inverse-proportional relationship between $\alpha$ and $\nu$ (figure A2).

2) Maximum longevity is an estimated indicator of longevity; there is empirically a proportional relationship between $\nu$ and $\omega$ (figure A3).

**5. Complementarity of longevity**

1) Phenomenon: "As $\alpha$ increases (with quadratic coefficient of shape parameter), $\nu$ (and $\omega$) decreases (with quadratic coefficient of shape parameter)."

2) Logic: "For the highest $S(t)$ (for longevity), $\beta(t)$ increases at $t<\alpha$ and $\beta(t)$ decreases at $t>\alpha$." In consequence of longevity, since $\beta_2$ indicates the decrease of $\beta(t)$ at $t>\alpha$, the quantity of $\beta_2$ increases with longevity (or $\alpha$). This simultaneously induces the decrease of $\nu$ (and $\omega$). There are obviously complementary aspects between $\alpha$ and $\nu$ (and $\omega$).

**6. Ultimate limit**

1) Possible limit: $\nu\rightarrow 0$ with longevity; $\beta_1\rightarrow 0$ or $\beta_2\rightarrow\infty$. There is no clue that $\beta_1$ approaches zero; instead, it is observed that $\beta_2$ (quadratic coefficient of shape parameter) increases as increasing $\alpha$.

2) Estimation: There seems to exist an ultimate limit of $\omega$ as $\beta_2$ increases. If we assume that $\omega$ has an exponential relationship with $\beta_2$, the result of regression analysis is given as follows: $\omega\approx 124.0+183.1\times\exp(\beta_2/0.0004)$ ($r^2=0.9395$). Therefore, the ultimate value can be estimated to be ~124 years for Sweden, Switzerland and Japan (figure A4).

**7. Implication**

"Human longevity is ultimately limited at an intrinsic (or biological) limit."

**Figures**

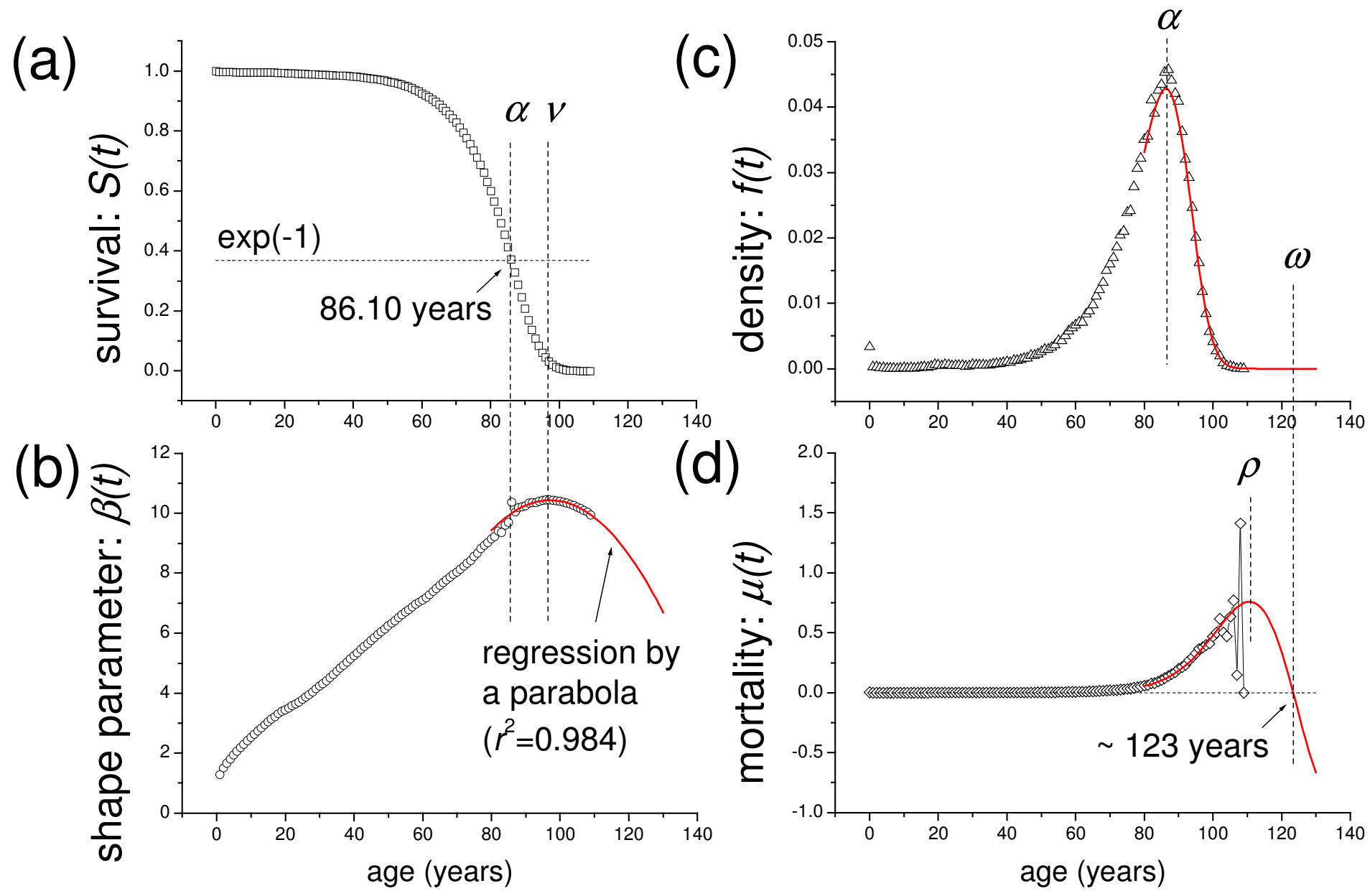

Figure A1. Demographic trajectories and singular points for Sweden (2002) [this figure was submitted to *Biology Letters*].

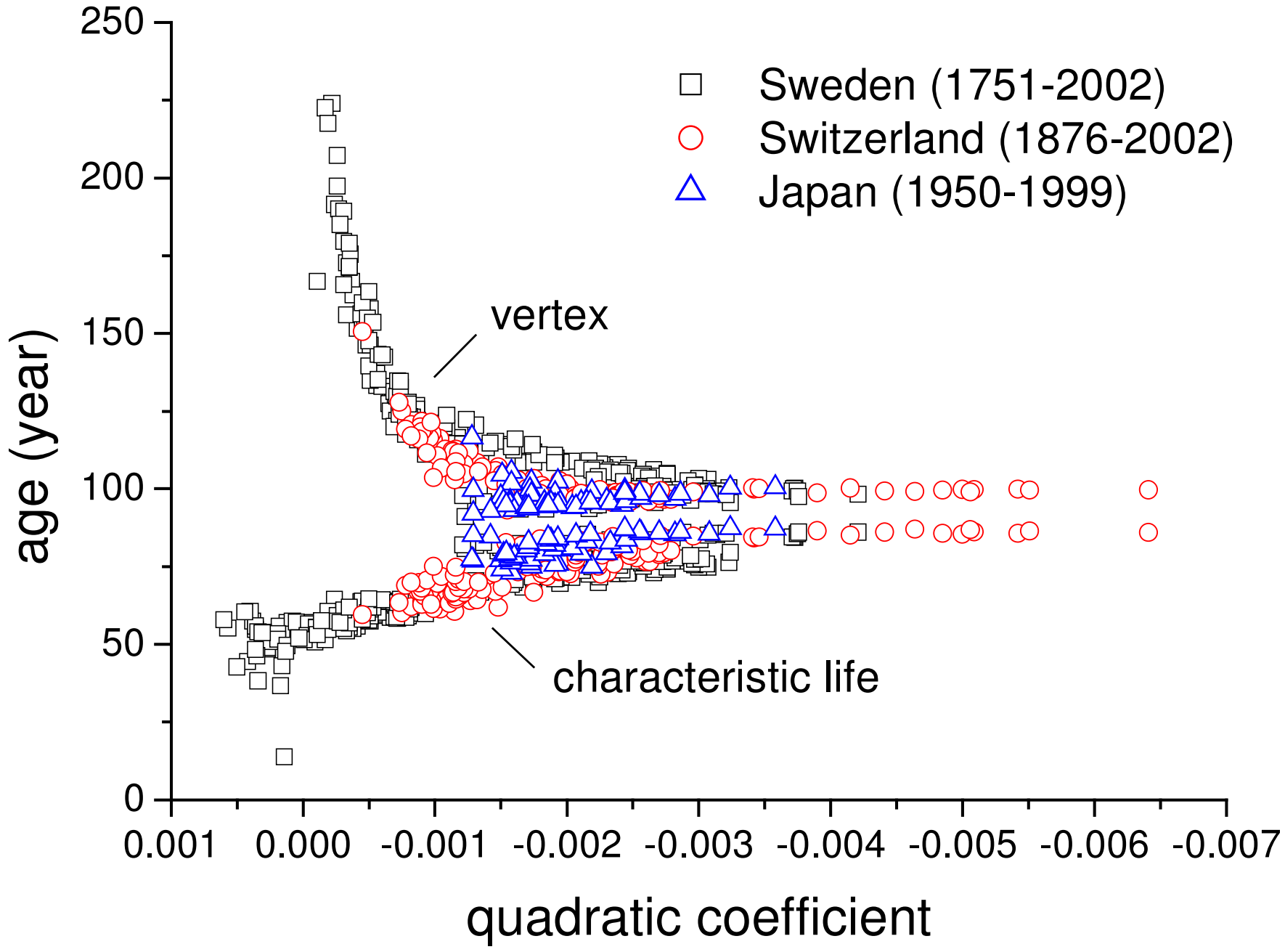

Figure A2. Complementarity of characteristic life and vertex of shape parameter.

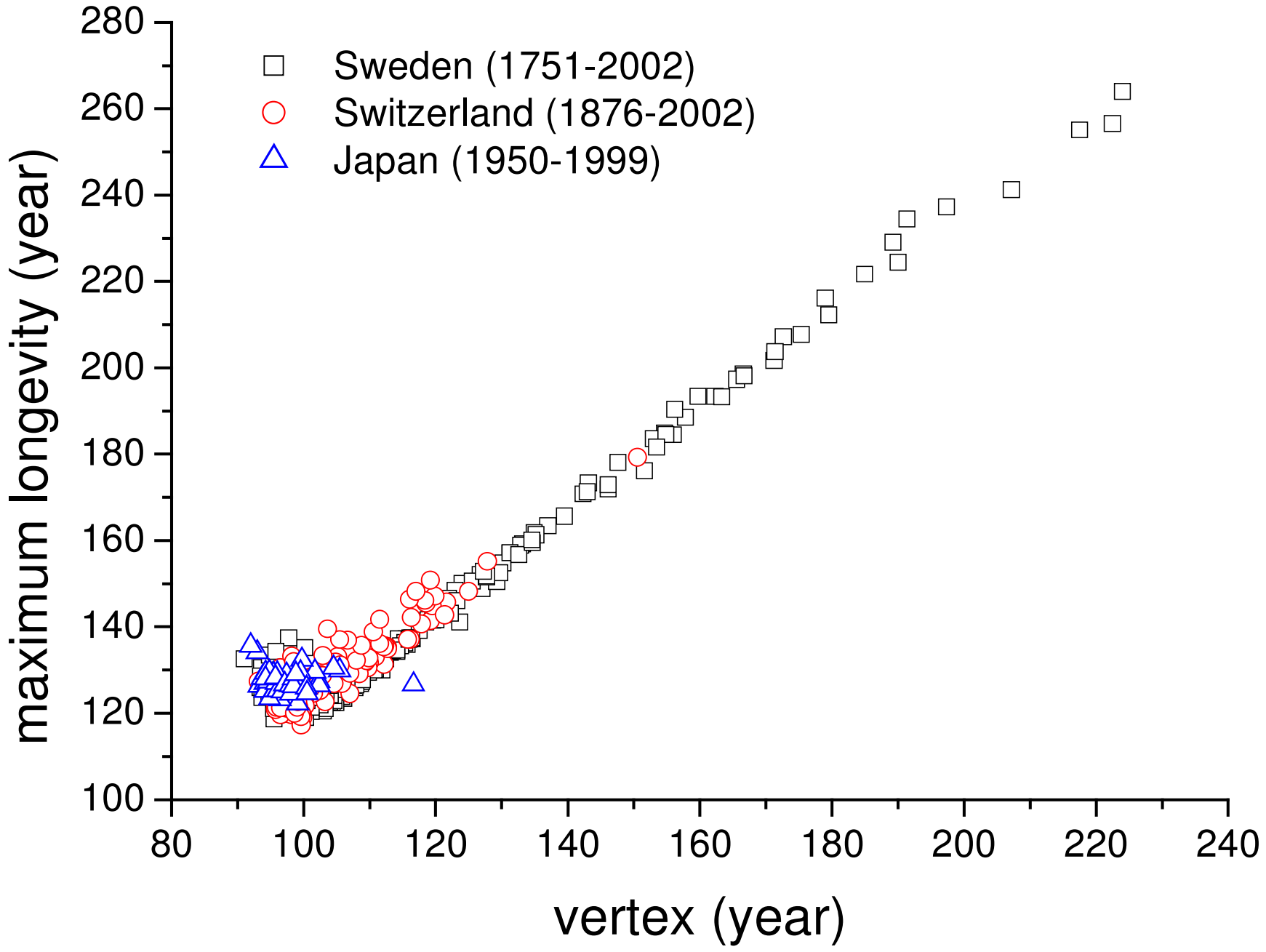

Figure A3. Relationship of maximum longevity and vertex of shape parameter.

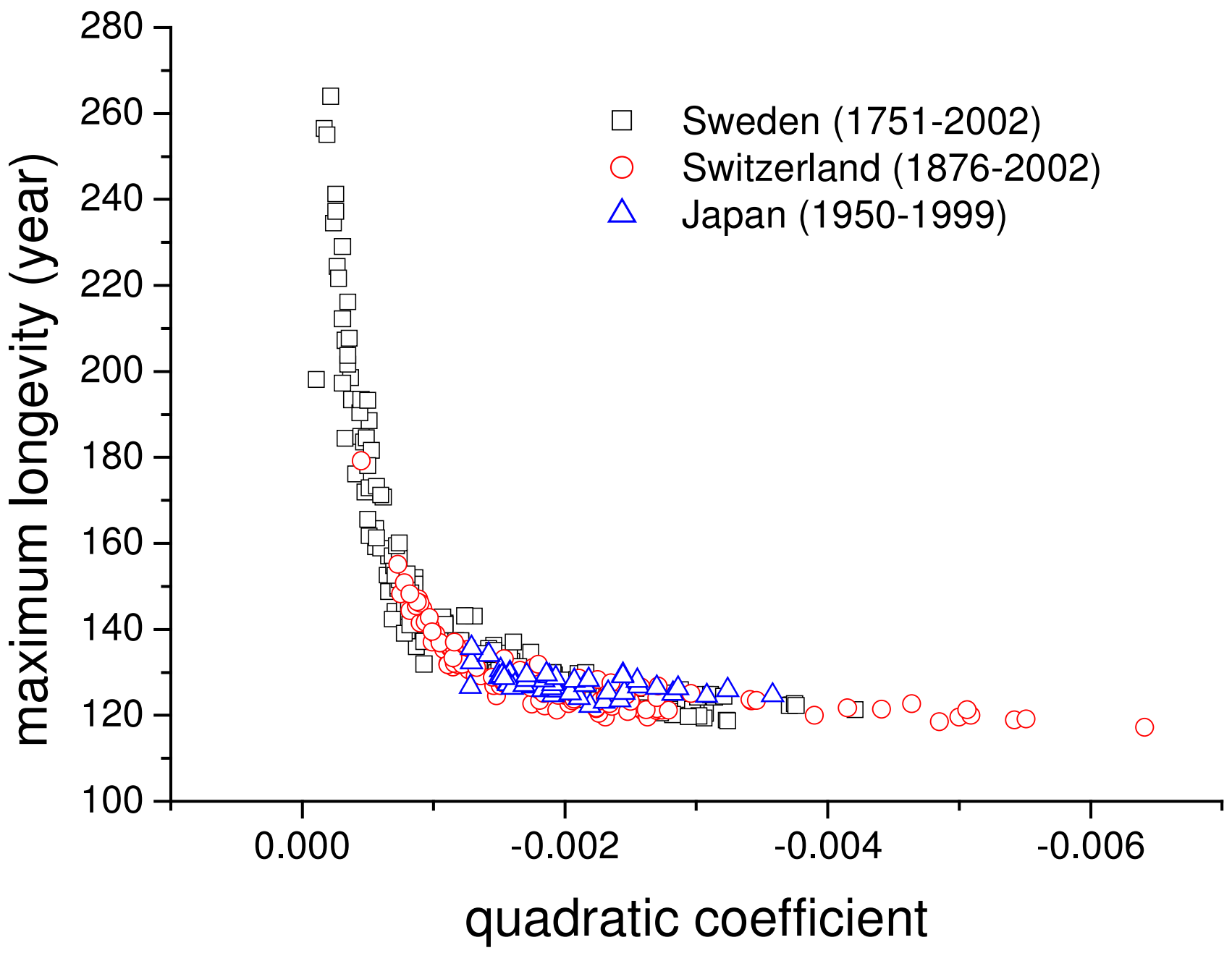

Figure A4. Trend of maximum longevity with quadratic coefficient of shape parameter (this figure was submitted to *Proceedings B*).